\documentclass[a4paper]{jpconf}

\usepackage{graphicx}
\usepackage{xspace}

\usepackage{aasjournals}
\usepackage{amsmath}
\usepackage{amssymb}
\usepackage[hidelinks]{hyperref}
\usepackage[sort&compress,numbers]{natbib}

\usepackage{color}
\newcommand{\isot}[2]{$^{#2}\mathrm{#1}$}
\newcommand{\isotm}[2]{{}^{#2}\mathrm{#1}}

\newcommand{\logten}[1]{\log_{10}{#1}}

\newcommand{\unitspace}{\ensuremath{\,}}
\newcommand{\usp}{\unitspace}

\newcommand{\unitstyle}[1]{\ensuremath{\mathrm{#1}}}
\newcommand{\power}[2]{\ensuremath{{#1}^{#2}}}

\newcommand{\centi}{\unitstyle{c}}

\newcommand{\Mega}{\unitstyle{M}}

\newcommand{\meter}{\unitstyle{m}}

\newcommand{\cm}{\centi\meter}
\newcommand{\gram}{\unitstyle{g}}

\newcommand{\grampercc}{\gram\usp\power{\cm}{-3}} 

\newcommand{\eV}{\unitstyle{eV}}        
\newcommand{\MeV}{\Mega\eV} 

\newcommand{\maestro}{{\sffamily MAESTROeX}}

\newcommand{\pynucastro}{{\sffamily pynucastro}}

\newcommand{\yt}{{\sffamily yt}}

\begin{document}

\title{Sensitivity of 3D Convective Urca Simulations to Changes in Urca Reactions}

\author{B~Boyd$^{1, 2}$, A~Smith Clark$^1$, A~C~Calder$^{1,2}$, D~M~Townsley$^3$, and M~Zingale$^1$}

\address{$^1$ Department of Physics and Astronomy, 
Stony Brook University, Stony Brook, NY 11794-3800, USA}
\address{$^2$ Institute for Advanced Computational Science,
Stony Brook University, Stony Brook, NY 11794-5250, USA}
\address{$^3$ Department of Physics and Astronomy, University of Alabama, Tuscaloosa, AL 35487-0324, USA}

\ead{boyd.brendan@stonybrook.edu}

\begin{abstract}
    A proposed setting for thermonuclear (Type Ia) supernovae is a white dwarf that has gained mass from a companion to the point of carbon ignition in the core. 
    There is a simmering phase in the early stages of burning that involves the formation and growth of a core convection zone. 
    One aspect of this phase is the convective Urca process, a linking of weak nuclear reactions to convection that may alter the composition and structure of the white dwarf. 
    Convective Urca is not well understood and requires 3D fluid simulations to realistically model. 
    Additionally, the convection is relatively slow (Mach number less than 0.005) so a low-Mach method is needed to make simulating computationally feasible. 
    Using the \maestro\ low-Mach hydrodynamics code, we investigate recent changes to how the weak reactions are modeled in the convective Urca simulations. 
    We present results that quantify the changes to the reaction rates and their impact on the evolution of the simulation. 
\end{abstract}

\section{Introduction}
\label{sec:intro}

Type Ia Supernovae (SNe Ia) are extremely bright thermonuclear explosions. SNe Ia light curves are standardizable, allowing them to be used as distance indicators, providing value to many areas of astronomy, particularly cosmology  \cite{phillips1993, perlmutter1999}. 
It is well accepted that SNe Ia arises from the thermonuclear explosion of white dwarf stars in a binary system. 
However, the specific progenitor system of SNe Ia is still an open question. 
One proposed setting is the single degenerate case in which a degenerate white dwarf accretes material from a companion star. 
As the white dwarf mass approaches the Chandrasekhar mass ($\sim 1.4 \; \mathrm{M_{\odot}}$), the core becomes hot and dense enough to start carbon burning. 
Initially, this burning is relatively slow and will last about 1,000 to 10,000 years prior to the final explosive event. 
This phase is called the carbon simmering phase, and it may impact the composition of the progenitor, which can alter the final nucleosynthesis observed in the SN Ia. 

During the simmering phase, the carbon burning releases enough energy to drive core convection in the white dwarf. 
This convection combines with weak nuclear reactions to produce local cooling due to neutrino emission, called the convective Urca process \cite{paczynski1972}.
To investigate this poorly understood mechanism, we need to model both the convection and weak reactions accurately.
Previous work \cite{stein-wheeler2006, lesaffre2005, piersanti2022} on the convective Urca process has largely made either 1D or 2D approximations to model the turbulent convection, an inherently 3D process. 
We use the low Mach hydrodynamic code \maestro\ to model convective Urca in 3D, building on an earlier study \cite{willcox2018, calder2019}.  
Additionally, we use weak reaction rates that depend on temperature and the electron density (which is important for this highly degenerate regime) \cite{suzuki2016}. 
In investigating the convective Urca process, slight changes have been made to how we estimate these weak reaction rates. 
We present the results of these changes and the sensitivity of our simulations to such changes. 

In section \ref{sec:convect-urca}, we describe the convective Urca process in greater detail. Next, in section \ref{sec:num-method}, we briefly describe our simulation setup. In section \ref{sec:changes}, we show the changes to the Urca reaction rates. In section \ref{sec:results}, we look at the impact of these changes on a test run of our simulation. Finally, in section \ref{sec:conclusions} we conclude with our takeaways from this analysis.

\section{The Convective Urca Process}
\label{sec:convect-urca}
 
The Urca process describes the combination of a beta decay and electron capture that link two isotopes, called an Urca pair. 
The reactions work as follows for a pair of nuclei with the same atomic mass number, $A$, and proton numbers, $Z-1$ and $Z$ respectively:
\begin{equation}
    \begin{split}
        \left(Z-1,A\right)           &\longrightarrow \: \left(Z,A\right) + e^{-} + \bar{\nu}_e  \\
        \left(Z,A\right) + e^{-} &\longrightarrow \: \left(Z-1,A\right) + \nu_e
    \end{split}
\end{equation}

The process provides some local cooling due to neutrino emission in each weak reaction. 
In a degenerate white dwarf, these weak reactions depend on temperature and the electron density. 
In most regions only one reaction will be active, i.e.\ electron captures at higher densities and beta decays at lower densities. 
The temperature dependence on the rate is relatively minor at the densities of interest, but it works to increase both the electron capture and beta decay reaction rates.
The transition between these regions, where the reaction rates are equal, is called the Urca shell. 
In a convection zone that encompasses the Urca shell, material cyclically rises above and falls below the shell. 
This allows a single nucleus to undergo repeated reactions and drive the neutrino cooling. 
The cyclic nature is key to convective Urca, enabling small abundances of the Urca pair to have a meaningful impact on the white dwarf's evolution.
In addition to neutrino cooling, the convective Urca process may impede the convective flow \cite{lesaffre2005, stein-wheeler2006}, but a conclusive 3D analysis is needed to better understand this effect. We look to investigate this idea in future work.

There are many Urca pairs relevant to simmering white dwarfs, but we focus this paper on the important $A=23$ Urca pair, \isot{Na}{23}-\isot{Ne}{23}. 
The $A=23$ pair has a relatively low $Q$-value, $Q = 4.38 \; \MeV$ \cite{suzuki2016}, and is relatively abundant in the white dwarf at the onset of carbon burning, $X(\isotm{Na}{23}) \approx 10^{-4}$ \cite{martinez-rodriguez2016, schwab2017a, piersanti2017}. 
The $A=23$ Urca shell is approximately located at a density $\rho_{\rm{urca}} \sim 1.7 \times 10^9$ \grampercc\  \citep{suzuki2016}. 

\section{Numerical Methods and Simulation Setup}
\label{sec:num-method}

We use the \maestro\ hydrodynamic code \citep{fan2019}, which is part of the AMReX-Astro suite of hydrodynamic codes for reactive astrophysical flows. 
\maestro\ is designed specifically to simulate stellar interiors and atmospheres in low Mach environments (i.e.\ when the fluid speed is slower than the sound speed). 
When modeling low Mach flow, a standard compressible finite-volume method is limited by the acoustic wave timescale as opposed to the advection timescale of the fluid. 
\maestro\ addresses this issue by using an elliptic constraint on the velocity field, derived from a low Mach approximation ($\mathrm{Ma} \lesssim 0.1$). From this velocity constraint, \maestro\ uses a projection method to effectively filter out any acoustic waves while still accounting for density fluctuations due to heating, stratification, etc. 
This approach allows \maestro\ to efficiently and accurately model slow moving flows like core convection. 
Nuclear reactions are coupled to the fluid equations via Strang-Splitting. 

Our presented simulations are full 3D of a near Chandrasekhar mass white dwarf with a convective core. 
We use a Cartesian grid with 3 levels of refinement (20, 10, and 5 km resolution). The levels of refinement are defined at specified density cutoffs and we ensure the entire convective core is resolved to the finest level. 
The outer regions of the white dwarf are only resolved to the point that hydrostatic equilibrium is maintained. Further resolution outside the convective core would only increase the computational expense without providing further insight or accuracy to the mechanisms in the core region.

\begin{figure}[ht]
    \centering
    \includegraphics[width=0.8\textwidth]{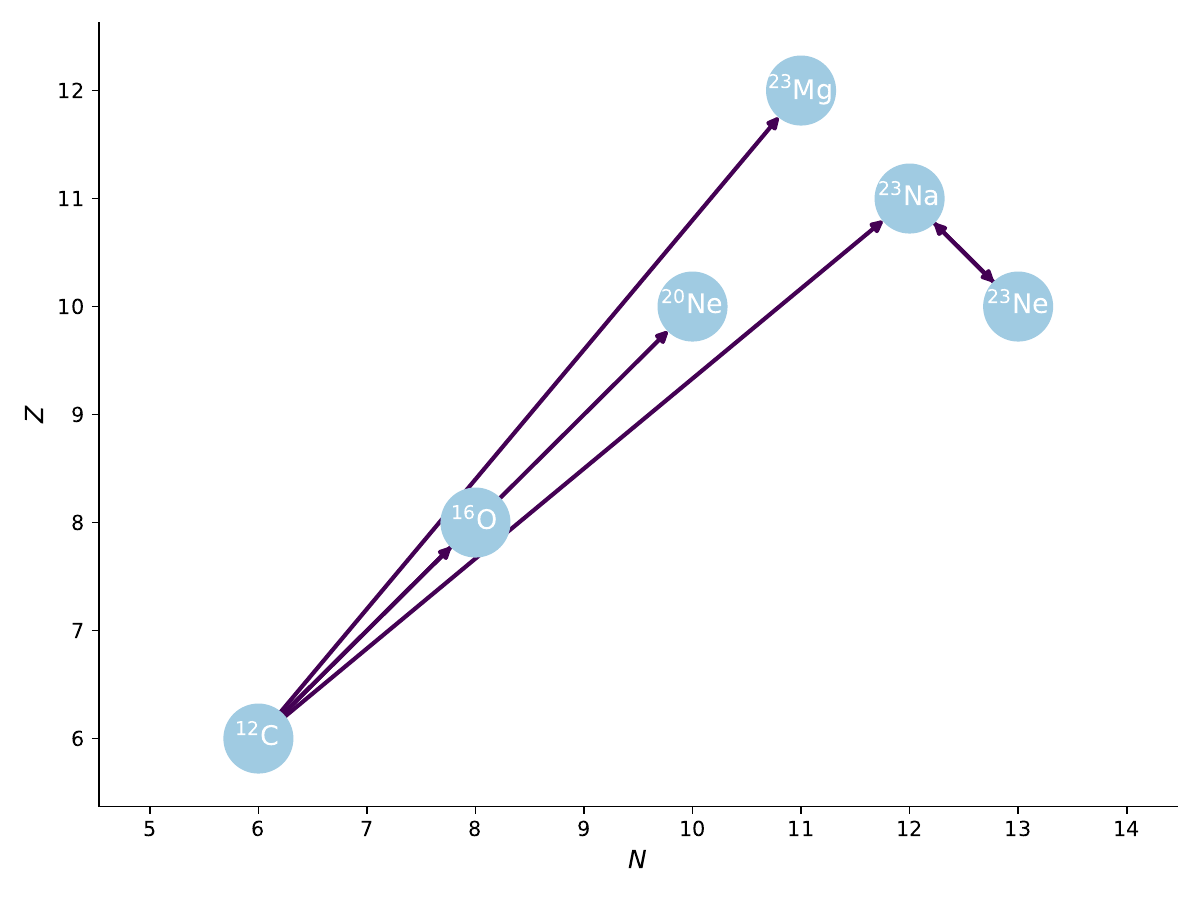}
    \caption{Graphical description of the reaction network. The x-axis displays the neutron number and the y-axis displays the proton number. Isotopes included in the network are labeled. The arrows indicate the direction of the reactions (double-sided arrows indicate both directions are included).}
    \label{fig:urca-net}
\end{figure}

The initial conditions are based on 1D stellar evolution models \cite{martinez-rodriguez2016} and were implemented in \maestro\ by a previous study \cite{willcox2018, calder2019}. 
We use a reaction network generated by the \pynucastro\ python package \cite{smith2023}. 
This network incorporates a simple carbon burning scheme and the $A=23$ Urca reactions, see Figure \ref{fig:urca-net}. 
Rates related to carbon burning come from JINA REACLIB \cite{cyburt2010}. 
We include the weak Urca reaction rates by interpolating a table of calculated values \cite{suzuki2016}. 

\begin{figure}[ht]
    \centering
    \includegraphics[width=0.8\textwidth]{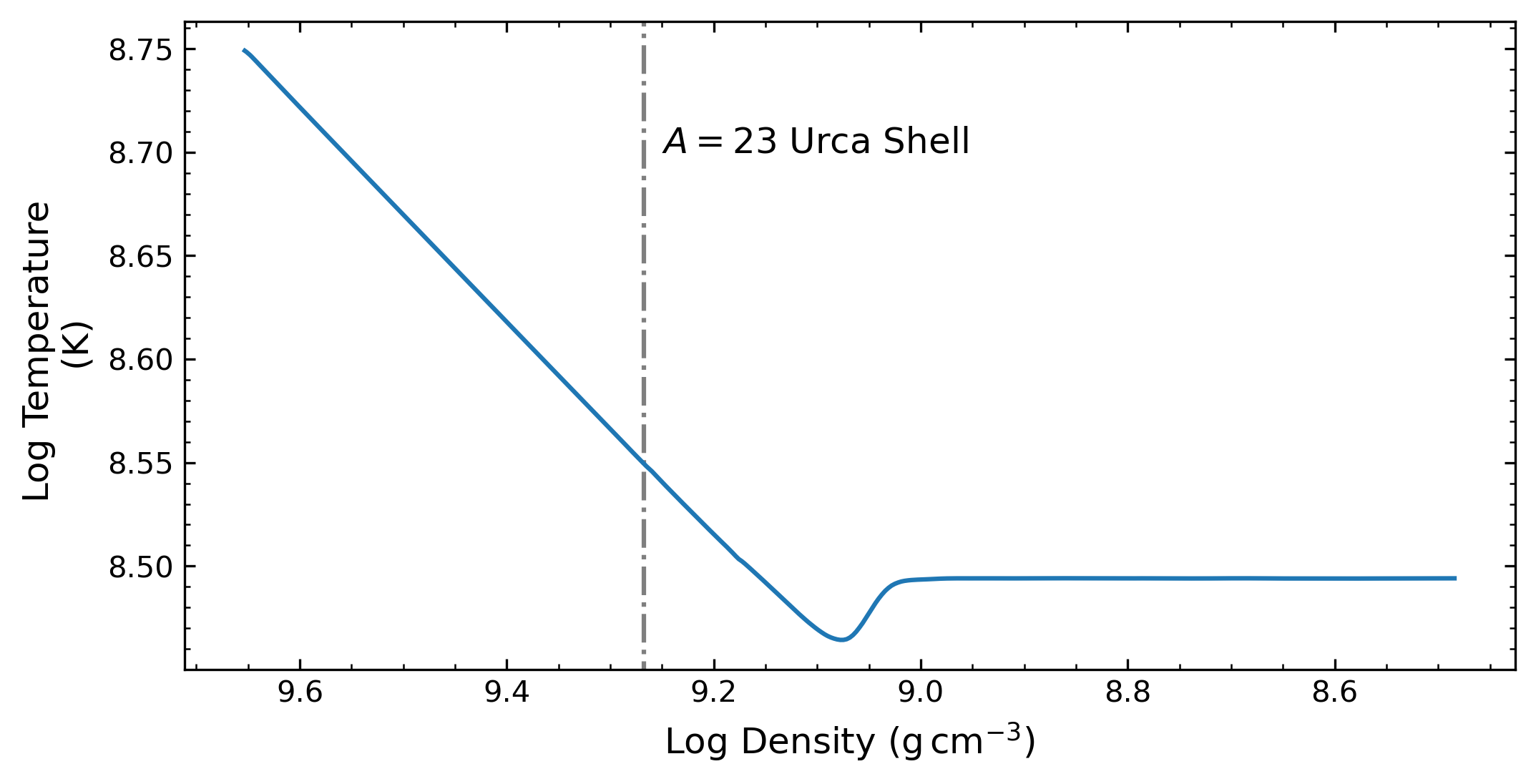}
    \caption{The profile of density vs temperature for the white dwarf star. The grey vertical line marks the location of the Urca shell.}
    \label{fig:T-rho}
\end{figure}

In our analysis, we use a simulation that has developed a fully convective core in a quasi-steady state. 
The convection zone has a typical Mach number of $\sim 10^{-3}$, which indicates the need for a low Mach number method like \maestro. 
The simulation has a central temperature of $5.5 \times 10^8 \; \mathrm{K}$ and a central density of $4.5 \times 10^9 \; \grampercc$. 
The temperature-density profile can be seen in Figure \ref{fig:T-rho}. 
This illustrates the isentropic core and the convectively stable isothermal envelope. 
Note the dip in the profile is due to the convection zone, and thus isentropic region, expanding outward from our initial state.
The white dwarf is primarily made of 40\% Carbon and 60\% Oxygen, with a trace amount of $A=23$ Urca pair ($\mathrm{X}(\isotm{Na}{23}) + \mathrm{X}(\isotm{Ne}{23}) \approx 10^{-4}$). 

\section{Changes to Urca Reactions}
\label{sec:changes}

The Urca reaction rates we use are tabulated over a range of temperatures, $\mathrm{T}$, and electron densities, $\rho Y_e$. 
We calculate the reaction rate for a given temperature and electron density by interpolating over this table. 
Prior to version 2.1 of \pynucastro\, this interpolation had been done strictly bi-linearly, i.e.\ interpolating between values of $\mathrm{T}$ and $\rho Y_e$.
However, the tables were constructed in log space and thus it's more accurate to interpolate in log space as well, i.e.\ interpolating between values of $\logten{\mathrm{T}}$ and $\logten{\rho Y_e}$. 
The shift from linear-space to log-space interpolation decreases the estimated reaction rate for values far from the table's spacing, $\Delta \logten{\mathrm{T}} = 0.05$ and $\Delta \logten{\rho Y_e} = 0.02$.
The significance of this shift depends on the temperature, density and composition. 

\begin{figure*}[!ht]
    \centering
    \includegraphics[width=\linewidth]{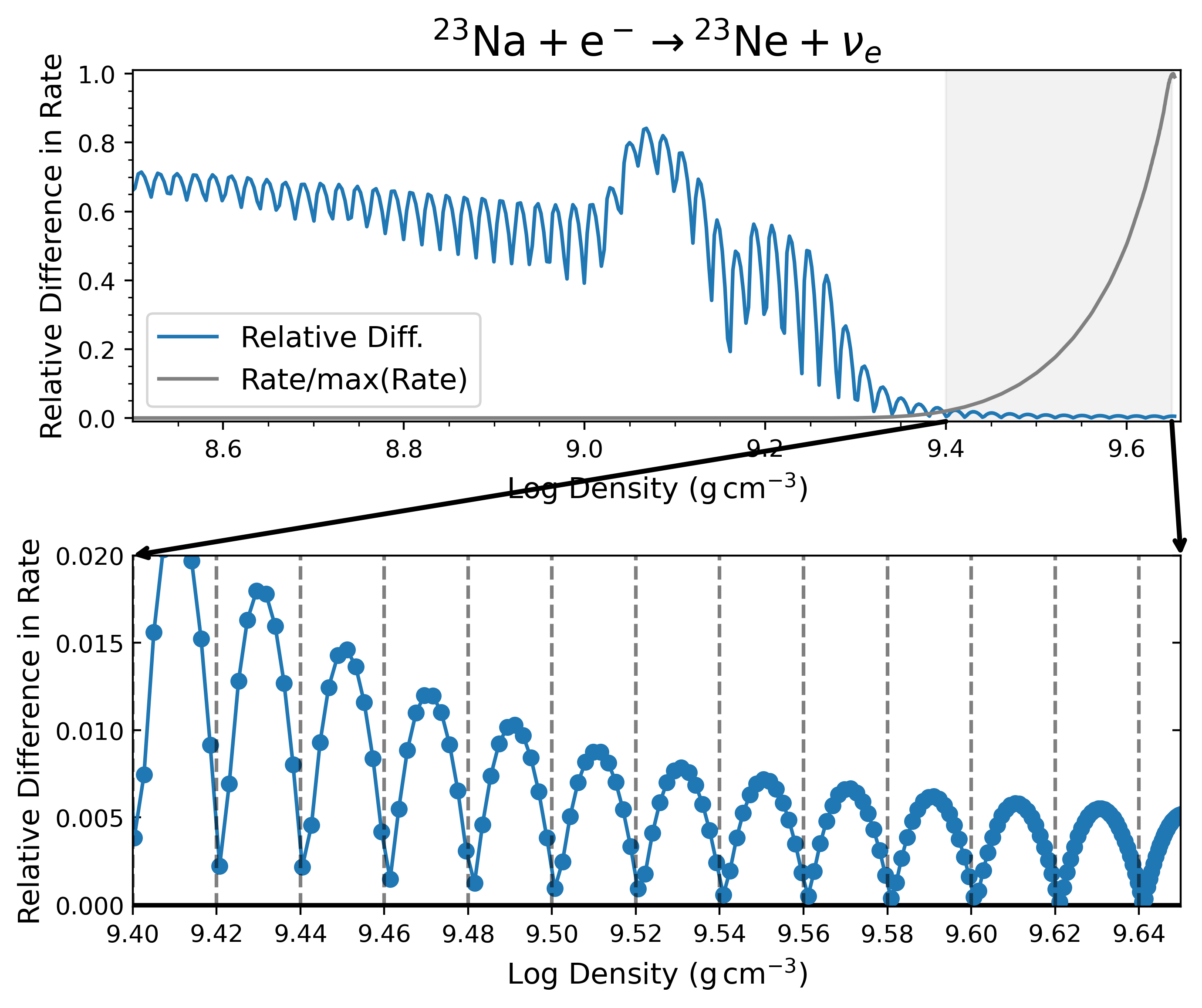}
    \caption{A Radially averaged profile of the electron capture reaction rates. 
    The bottom plot is a zoomed in version of the shaded region in the top plot.
    In each plot, the blue curve is the relative difference between the Linear-space and Log-space interpolation schemes for the initial state of the simulation (see Figure \ref{fig:T-rho}). 
    In the top plot, the grey curve is a normalized profile of the reaction rate. 
    In the bottom plot, the grey vertical dashed lines mark the spacing of the tabulated values.}
    \label{fig:diff-ecap}
\end{figure*}

\begin{figure*}
    \centering
    \includegraphics[width=\linewidth]{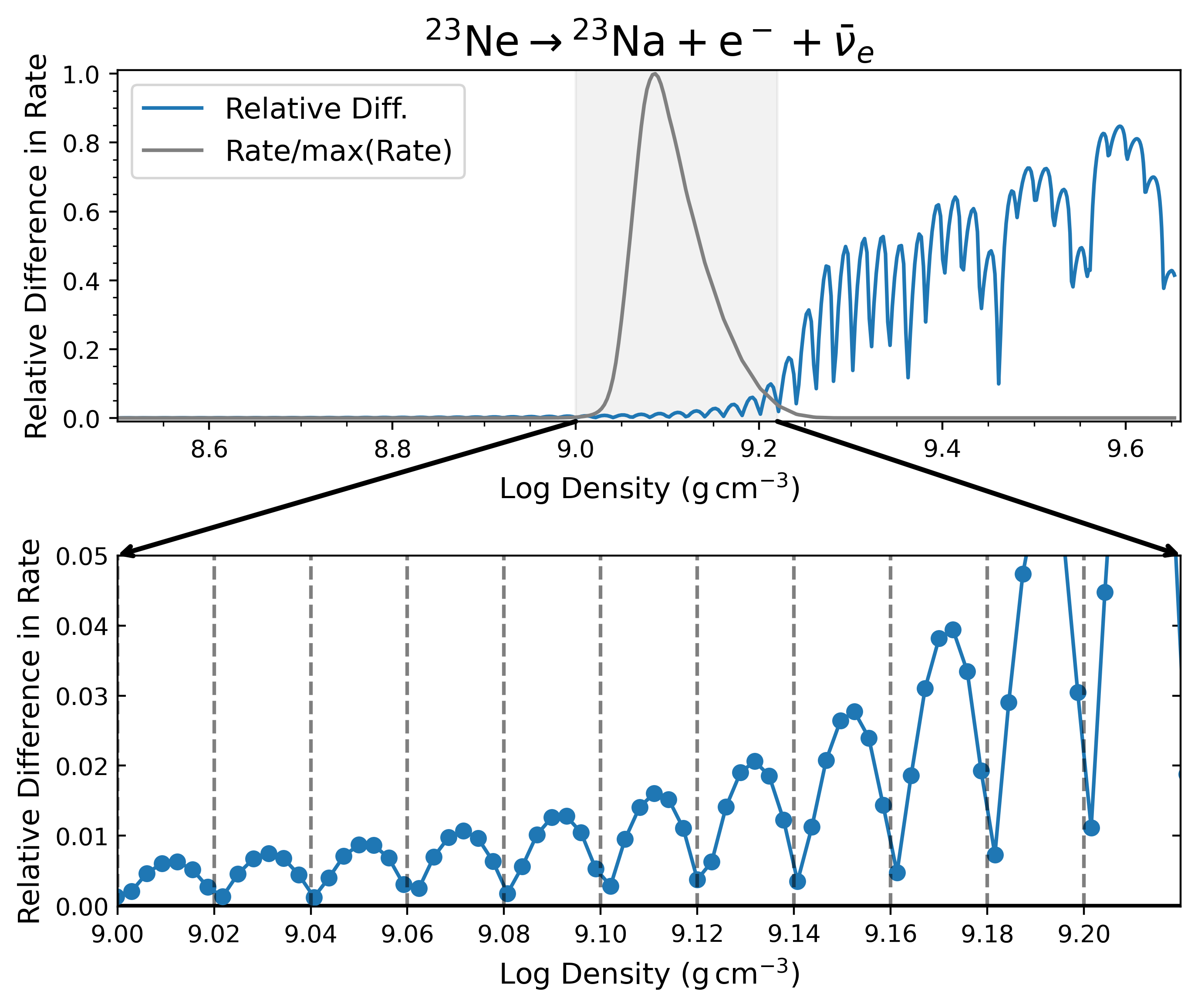}
    \caption{A Radially averaged profile of the beta decay reaction rates. 
    The bottom plot is a zoomed in version of the shaded region in the top plot.
    In each plot, the blue curve is the relative difference between the Linear-space and Log-space interpolation schemes for the initial state of the simulation (see Figure \ref{fig:T-rho}). 
    In the top plot, the grey curve is a normalized profile of the reaction rate. 
    In the bottom plot, the grey vertical dashed lines mark the spacing of the tabulated values.}
    \label{fig:diff-beta}
\end{figure*}

To quantify the change in the Urca rates for the conditions in our simulation, we generate a radial profile of the white dwarf and calculate the average reaction rate at a given radius for each interpolation scheme. 
These Urca reactions primarily depend on the density, so we plot the difference in rates vs density (see Figure \ref{fig:diff-ecap} and \ref{fig:diff-beta}).
The two figures suggest the difference in rates can be large, up to 80\%, but only in the regions where the rates are relatively small and the reaction is negligible (see solid grey lines in Figure \ref{fig:diff-ecap} and \ref{fig:diff-beta}). 
Looking at the zoomed in plots in Figure \ref{fig:diff-ecap} and \ref{fig:diff-beta}, we see that the difference between interpolation scheme is upwards of 5\% in these regions where the reaction is active. 
The vertical, dashed lines indicate the density spacing of the tabulated calculations which is why the difference approaches zero for these densities.

\section{Results}
\label{sec:results}

We investigated the sensitivity of our simulation to the change in the Urca rates by comparing two runs with different interpolation schemes. 
The initial state, described in Section \ref{sec:num-method}, is from a simulation that had settled to a quasi-steady state, using the linear-space interpolation scheme. 
This allows us to start with a realistic velocity field and distribution of the Urca pair. 
We ran each simulation for 300 seconds of simulation time (about 4000 timesteps), which accounts for several convective turnovers. 

\begin{figure}[ht]
    \centering
    \includegraphics[width=0.7\textwidth]{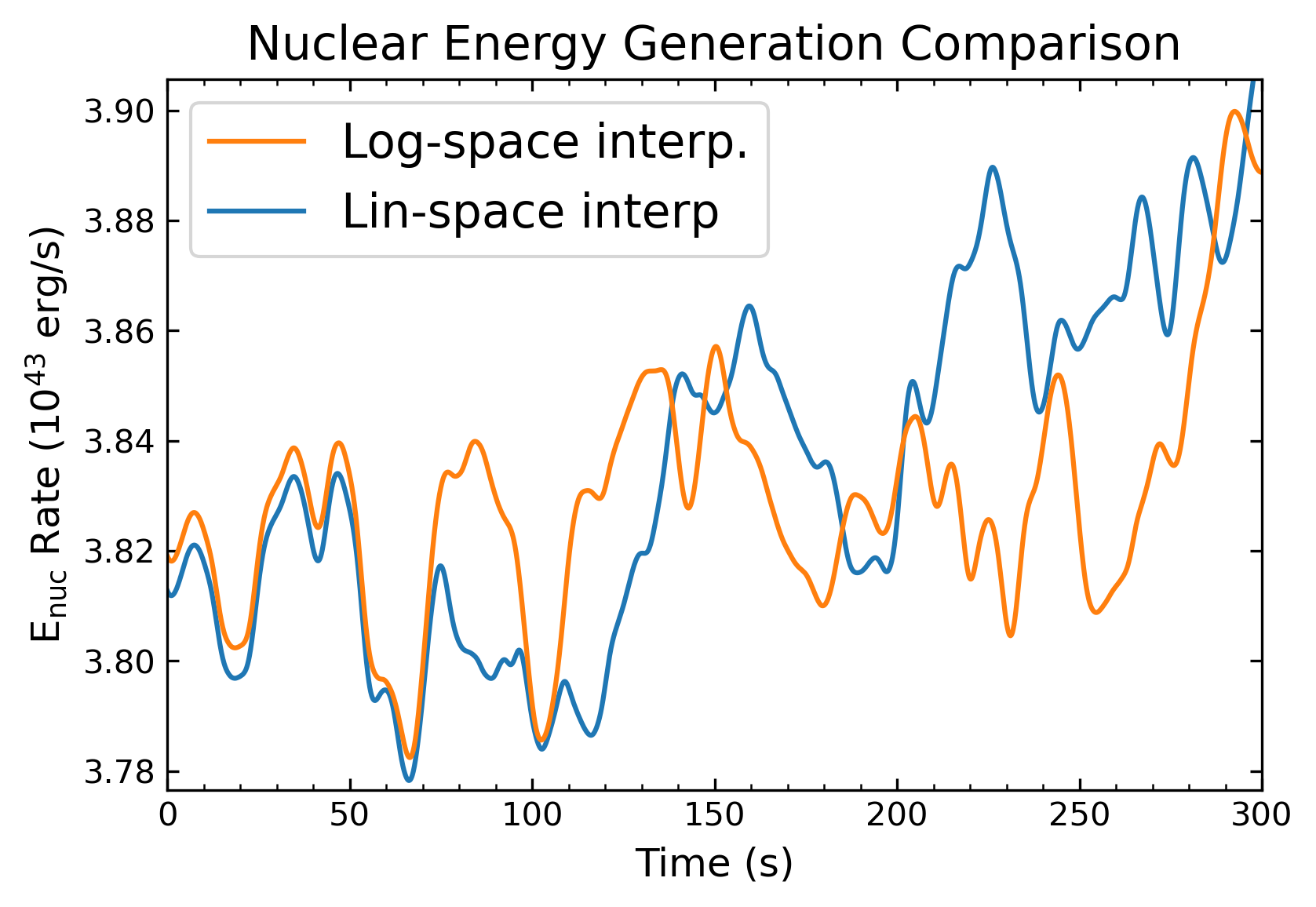}
    \includegraphics[width=0.7\textwidth]{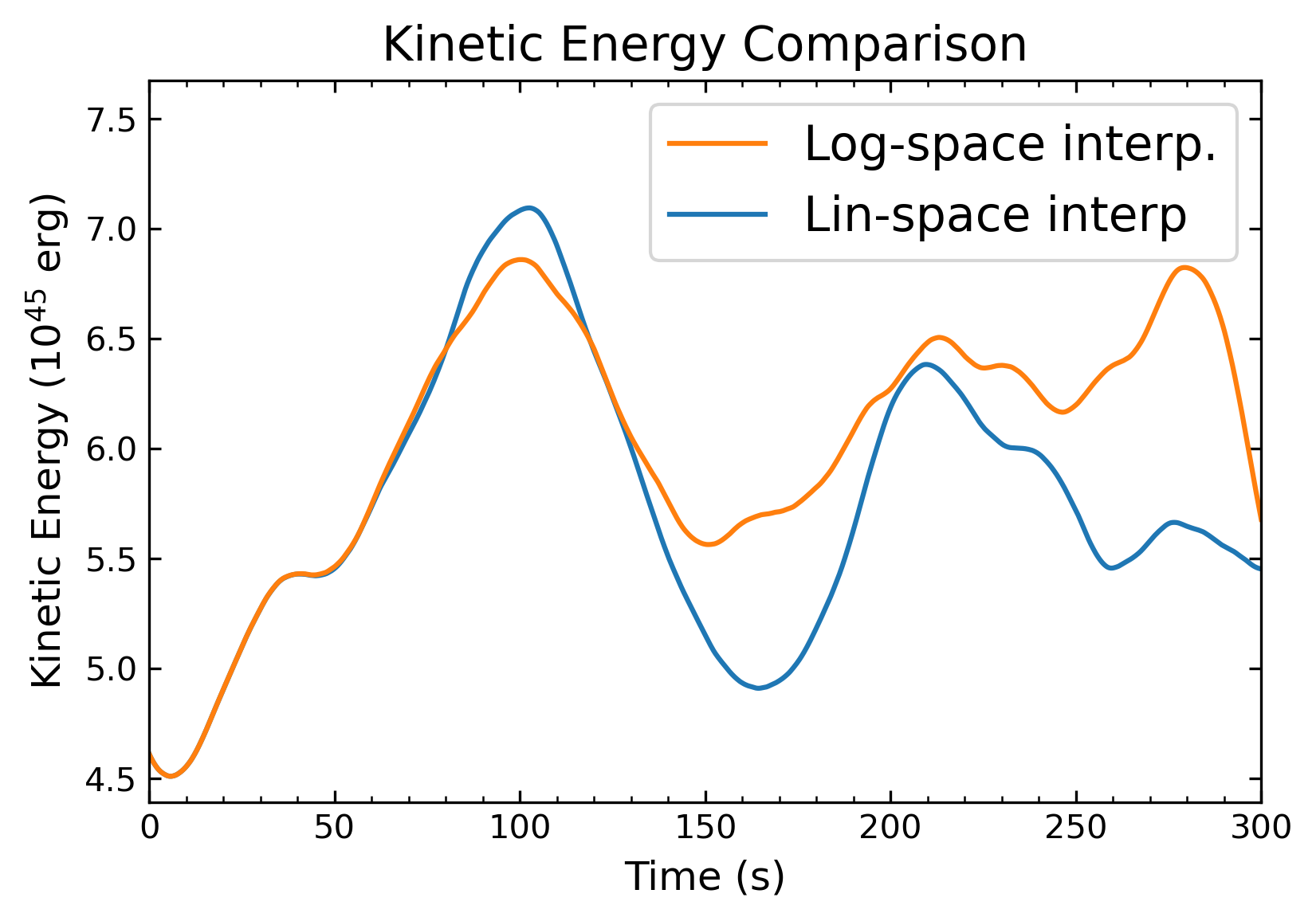}
    \caption{The top plot displays the time evolution of the nuclear energy generation rate integrated over the whole star. The bottom plot displays the time evolution of the total kinetic energy of the whole star. The blue curves represent the simulation using the linear-space interpolation scheme. The orange curves represent the simulation using the log-space interpolation scheme.}
    \label{fig:energy-gen}
\end{figure}

In Figure \ref{fig:energy-gen}, we plot the time evolution of the nuclear energy generation rate and the kinetic energy of the white dwarf. 
We calculate the specific energy generation rate by integrating the reaction network (see Figure \ref{fig:urca-net}) using an implementation of the VODE integrator \citep{brown1989}. 
The total nuclear energy generation rate for the star is then the mass integrated sum of the specific energy generation rate. 
The energy is primarily generated by the core carbon burning, but changes to the Urca reactions can easily be seen as well. 
For the first 70 seconds, the log-space interpolation has slightly higher energy generation due to fewer Urca reactions (i.e.\ less neutrino cooling). 
After about 70 seconds, the two simulations diverge in the energy being generated. 
The total kinetic energy is calculated by summing the kinetic energy of each cell in the grid. 
The evolution of this energy follows a similar pattern to the nuclear energy generation rate (i.e.\ the simulations diverge after $\sim$70 seconds). 
These differences indicates the changes to the Urca reactions have had an impact on the evolution of the simulation.

To further quantify the impact on the velocity field, we calculate the primary direction of the flow, i.e.\ the dipole moment. 
This is done by taking the density weighted average of the radial velocity in the x, y, and z directions as shown:

\begin{equation}
    \langle v_r \rangle_{x} = \left( \sum  \rho v_r \frac{x - x_c}{r} \right) / \sum \rho
\end{equation}

Here we are summing over each cell in the simulation. $x_c$ is the x coordinate for the center of the star and $r$ is the radial coordinate from the center. 
$\langle v_r \rangle_{y}$ and $\langle v_r \rangle_{z}$ are calculated analogously. 
Using these values, we calculate the angle $\theta$ from the z-axis and the azimuthal angle $\phi$ (in the x-y plane) as follows:

\begin{equation}
    \theta = \mathrm{arctan} \left( \frac{\sqrt{{\langle v_r \rangle_{x}}^2 + {\langle v_r \rangle_{y}}^2}}{\langle v_r \rangle_{z}} \right)
\end{equation}

\begin{equation}
    \phi = \mathrm{arctan} \left( \frac{\langle v_r \rangle_{y}}{\langle v_r \rangle_{x}} \right)
\end{equation}

The results are shown in Table \ref{tab:dipole-angle}. 
Although the general direction is similar (downward in the negative z direction), there is clear deviations in the net flow between the two simulations.

\begin{table}[ht]
    \centering
    \begin{tabular}{|c|c|c|} \hline 
       \hfill  & $\theta$ & $\phi$ \\ \hline 
       Initial State  & 167$^\circ$ & 348$^\circ$ \\ \hline 
       Log-space Interp.  & 168$^\circ$ & 7$^\circ$   \\ \hline 
       Lin-space Interp.   & 174$^\circ$ & 49$^\circ$ \\ \hline
    \end{tabular}
    \caption{Angular direction of the velocity field's dipole moment. $\theta$ is the polar angle (from the z-axis) ranging from 0 to 180 degrees. $\phi$ is the azimuthal angle (in the x-y plane) ranging from 0 to 360 degrees. The initial state is included for reference.}
    \label{tab:dipole-angle}
\end{table}

\begin{figure*}[ht]
    \centering
    \includegraphics[width=0.7\linewidth]{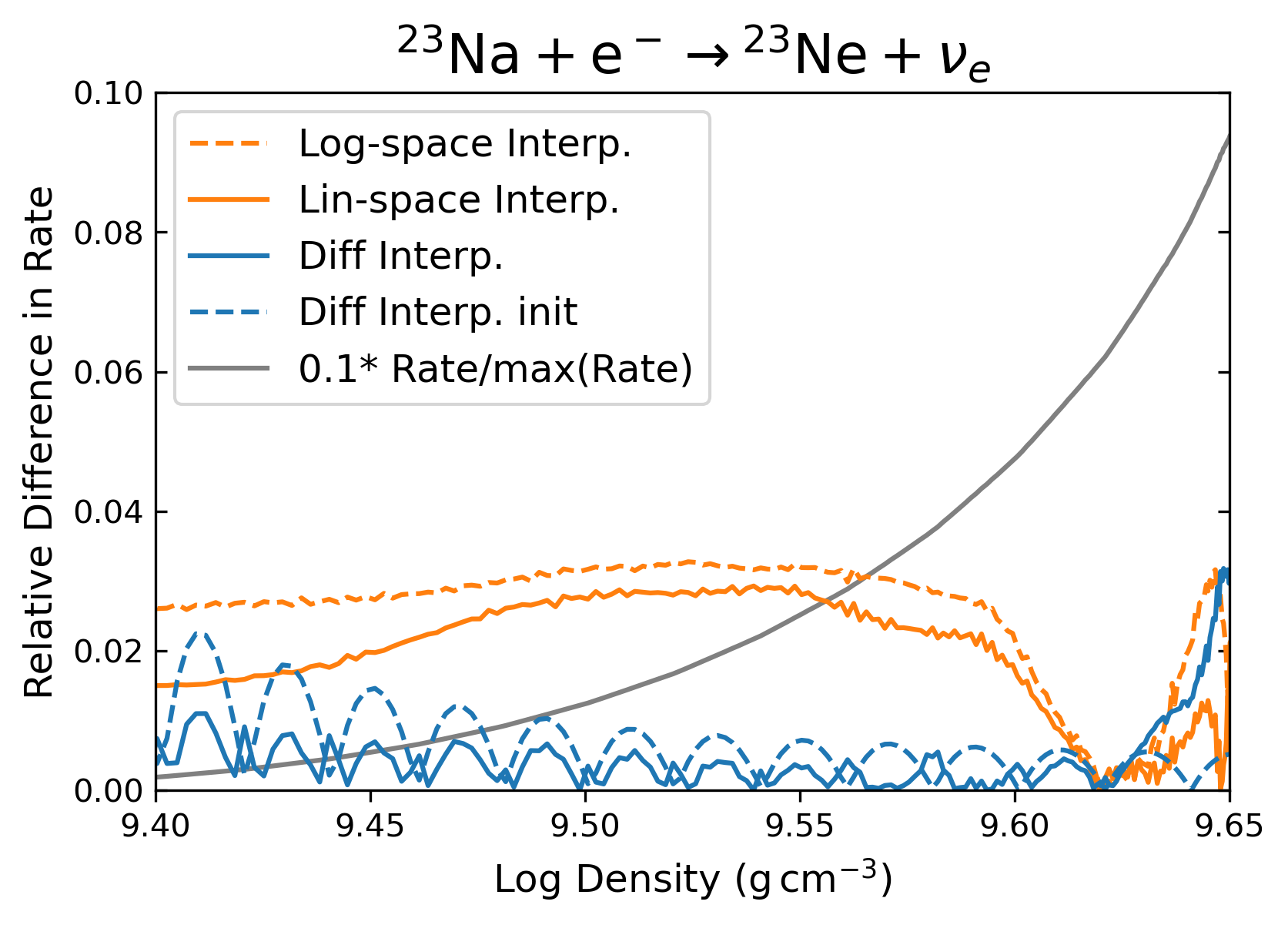}
  \caption{\label{fig:aftersim-ecap} The radially averaged profile of the electron capture reaction rate. The blue curves are the relative difference between the Log-space and Linear-space interpolations at the initial state (dashed) and after 300 seconds of evolution (solid). The orange curves are the relative differences between the initial state and the end state (after 300 seconds) for a given interpolation scheme (Log-space dashed, Lin-space solid). The grey curve is the normalized rate profile for reference.}
\end{figure*}

\begin{figure*}[ht]
    \centering
    \includegraphics[width=0.7\linewidth]{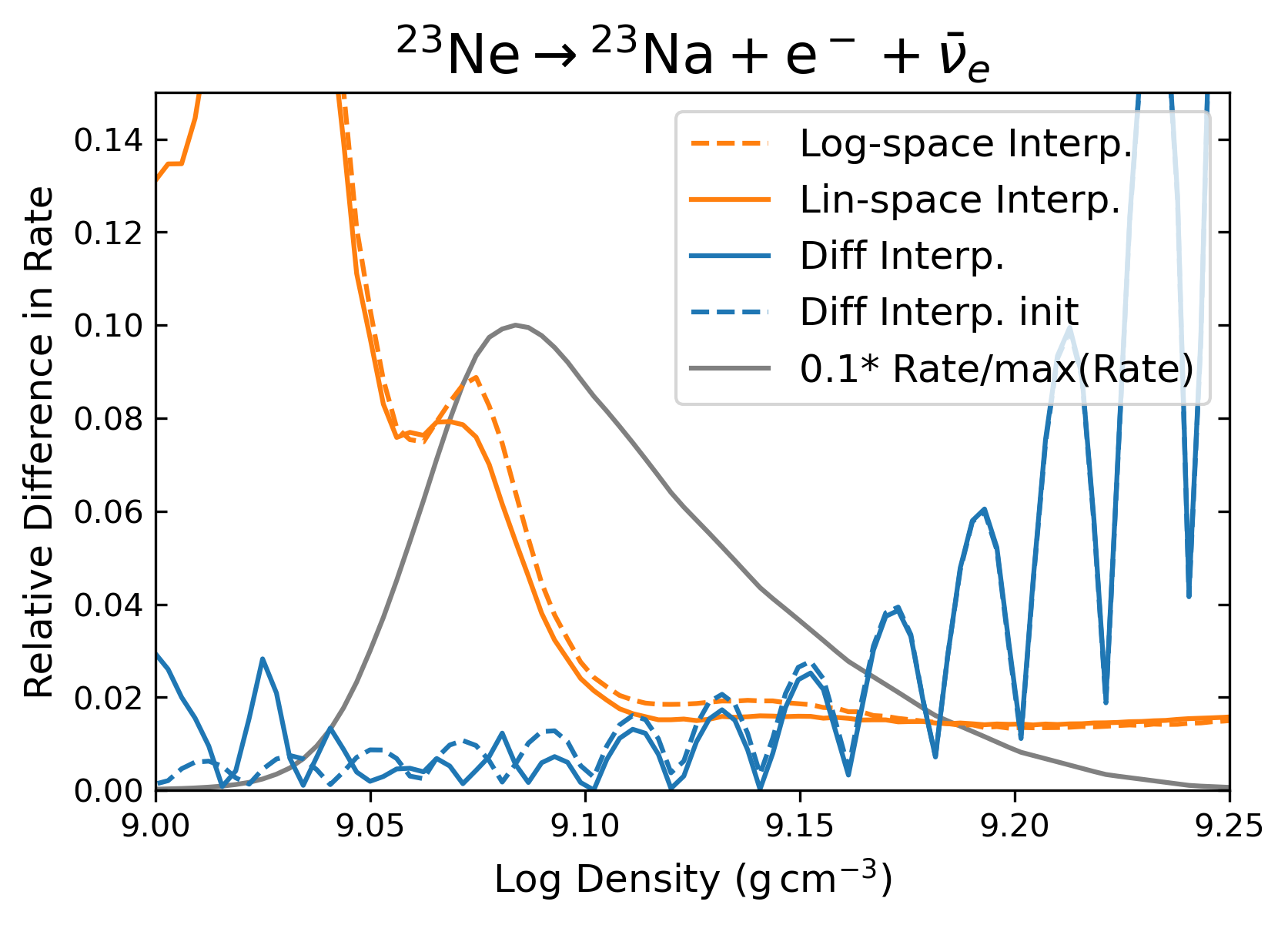}
  \caption{\label{fig:aftersim-beta} The radially averaged profile of the beta decay reaction rate. The blue curves are the relative difference between the Log-space and Linear-space interpolations at the initial state (dashed) and after 300 seconds of evolution (solid). The orange curves are the relative differences between the initial state and the end state (after 300 seconds) for a given interpolation scheme (Log-space dashed, Lin-space solid). The grey curve is the normalized rate profile for reference.}

\end{figure*}

From the changes in energy generation rate, total kinetic energy and the directional dependence of the flow, it is clear that the small changes in the Urca reaction rates can alter a simulation. 
However, these simulations are very chaotic (i.e.\ sensitive to initial conditions) due to the highly non-linear system of equations we are solving.
This chaotic nature means the new interpolation changes may alter the simulation without having a significant impact on the conclusions we draw from averaged values over many convective turnovers. 
We compare the changes due to interpolation relative to the natural evolution of the simulation, see Figure \ref{fig:aftersim-ecap} and \ref{fig:aftersim-beta}. 
In our regions of interest, the reaction rates tend to vary more significantly in time (orange curves) than by the change in interpolation scheme (blue curves). 
This indicates the dynamic nature of the simulation leads to comparable or larger shifts in the rates over just a few convective turnovers, reducing the relative impact the new interpolation scheme has on the reaction rates over a long periods of time. 
The exception to this is the region from $\logten{\rho}$ 9.15 to 9.20 in Figure \ref{fig:aftersim-beta}. 
Here, the difference in interpolation scheme (blue curves) is larger and the reaction rate is still significant (10-20\% the peak value).


\section{Conclusions}
\label{sec:conclusions}

The convective Urca process is an interesting mechanism to provide cooling in a simmering white dwarf star. 
The reaction rates for these Urca reactions need to be tabulated and recent changes to \pynucastro\ (v2.1) has altered how the tables are interpolated. 
In the regions where the rates are most active, we found differences of up to 5\%. 
We investigate the sensitivity of our 3D simulations to these changes by comparing two simulations that had run for 300 s, a few convective turnovers.
We found that these two simulations deviated in their energy generation rate, total kinetic energy and in the general direction of the flow.
However, we account much of this to the highly non-linear nature of our equations. 
This makes the simulation particularly sensitive to even slight changes to initial conditions.
We found that the reaction rates largely varied in time more so than they varied in interpolation scheme for our regions of interest. 
We interpret this finding to mean that although the new interpolation scheme is more accurate, it does not significantly alter the quasi-steady state of the simulation and the time averaged values that we are interested in. 
Further time evolving these simulations, through more convective turnovers, may uncover more subtle effects which alter the simulation on longer timescales. With particular interest in the region above the Urca shell where differences in interpolation of the beta-decay reaction are most significant.
In future work, we look to further investigating the sensitivity of the convective Urca process to changes in the reaction network, including adding a more accurate prescription of neutron decay \cite{langanke2001}, using bi-cubic interpolation, adding additional Urca pairs, and using a larger carbon burning network.

\ack
\maestro\ is freely available on GitHub (\url{https://github.com/AMReX-Astro/}), and all problem setup files for the calculations presented here are in the code repository.  
This research was supported in part by the US Department of Energy (DOE) under grant DE-FG02-87ER40317
The reaction networks were generated using the \pynucastro\ library \cite{smith2023}.
This research used resources of the National Energy Research Scientific Computing Center, which is supported by the Office of Science of the U.S. Department of Energy under Contract No. DE-AC02-05CH11231. 
Visualizations and part of this analysis made use of \yt\ \cite{turk2011}. 

\bibliography{mybib}

\end{document}